\documentclass[superscriptaddress,aps,prd,twocolumn]{revtex4-2}
\usepackage[english]{babel}
\usepackage[utf8]{inputenc}
\usepackage{float}
\usepackage{amsmath,bm}
\usepackage{amssymb}
\usepackage{graphicx}
\usepackage{color}

\usepackage{hyperref}

\usepackage[normalem]{ulem}

\newcommand\mathplus{+}

\newcommand{\pr}[1]{\ensuremath{\left[#1\right]}}
\newcommand{\pc}[1]{\ensuremath{\left(#1\right)}}

\begin{document}
	
\title{Extracting nuclear matter properties from the neutron star matter equation of state using deep neural networks}

\author{Márcio Ferreira}
\email{marcio.ferreira@uc.pt}
\affiliation{CFisUC, 
	Department of Physics, University of Coimbra, P-3004 - 516  Coimbra, Portugal}
	
\author{Valéria Carvalho}
\affiliation{CFisUC, 
	Department of Physics, University of Coimbra, P-3004 - 516  Coimbra, Portugal}
	
\author{Constança Providência}
\email{cp@uc.pt}
\affiliation{CFisUC, 
	Department of Physics, University of Coimbra, P-3004 - 516  Coimbra, Portugal}

\date{\today}

\begin{abstract}
The extraction of the nuclear matter properties from neutron star (NS) observations is nowadays an important issue, in particular, the properties that characterize the symmetry energy which are essential to describe correctly asymmetric nuclear matter.  We use deep neural networks (DNNs) to map the relation between cold $\beta$-equilibrium NS matter and the nuclear matter properties. Assuming a quadratic dependence on the isospin asymmetry for the energy per particle of homogeneous nuclear matter and  using a Taylor expansion up to fourth order in the iso-scalar and iso-vector contributions, we generate a dataset of different realizations of $\beta$-equilibrium NS matter and the corresponding nuclear matter properties. The DNN model was successfully trained, attaining great accuracy in the test set. Finally, a real case scenario was used to test the DNN model, where a set of 33 nuclear models, obtained within a relativistic mean field approach or a Skyrme force description, were fed into the DNN model and the corresponding nuclear matter parameters recovered with considerable accuracy, in particular, the standard deviations $\sigma(L_{\text{sym}})= 12.85$ MeV and $\sigma(K_{\text{\text{sat}}})= 41.02$ MeV were obtained, respectively, for the slope of the symmetry energy and the nuclear matter incompressibility at saturation.
\end{abstract}

\maketitle

\section{Introduction}
The equation of state (EoS) of dense and asymmetric nuclear matter is still weakly known. Terrestrial conditions in the laboratory do not cover neither very high densities nor very large proton-neutron asymmetries. Neutron stars (NSs) constitute  the perfect environment where both dense  nuclear matter and very asymmetric nuclear matter exist \cite{glendenning97,Lattimer:2000nx,OertelRMP16,Rezzolla:2018jee,Burgio:2021vgk,Lattimer:2021emm}. 

The mass of several two solar mass pulsars has been determined with a quite small uncertainty in the last decade,  PSR~J1614-2230   \cite{Demorest2010,Fonseca2016,Arzoumanian2017} with  $M = 1.908 \pm~ 0.016 M_{\odot}$, PSR~ J0348 - 0432 with $M = 2.01 \pm~ 0.04~ M_{\odot}$ \cite{Antoniadis2013},  PSR J0740+6620 with $M = 2.08 \pm~ 0.07~ M_{\odot}$  \cite{Fonseca:2021wxt} and, very recently, J1810+1714 with $M = 2.13 \pm~ 0.04~ M_{\odot}$  \cite{Romani:2021xmb}. These observations put strong constraints on the high density EoS, and,  in particular, on the possible  onset of non-nucleonic degrees of freedom in the NS core, such as hyperons, quarks, or kaon and pion condensates. Also the detection of gravitational waves (GW) from binary neutron star mergers such as  the  LIGO/Virgo observations GW170817 \cite{Abbott:2018wiz} and GW190425 \cite{Abbott:2020khf}, together with the recent NICER (Neutron star Interior Composition ExploreR)  x-ray determination of the mass and radius of the pulsar PSR J0030+045  \cite{Riley_2019,Miller19} and  the determination of the radius of the pulsar PSR J0740+6620 from  NICER and XMM-Newton data \cite{Riley2021,Miller2021,Raaijmakers2021} set strong constraints on the NS EoS. Further  observations of NICER together with the future observations  programmed for the { enhanced X-ray Timing and Polarimetry mission} (eXTP) \cite{eXTP,eXTP:2018anb} and the { Spectroscopic Time-Resolving Observatory for Broadband Energy X-rays} (STROBE-X) \cite{STROBE-X} will allow the determination of the radius and mass of a large number of NSs with an uncertainty of only a few \%. Also, the largest radio  telescope Square Kilometer Array \citep{SKA} will be operating in the near future and the detection  of a large number of NSs, an order of magnitude larger than the presently known NSs,  is expected.\\
 
The knowledge of a large enough number of NS  will allow the determination of the $M(R)$ curve, which can be converted through different methods into the $\beta$-equilibrium EoS \cite{Lindblom1993,Lindblom:2010bb,Steiner:2010fz,Fujimoto2019}. However,  the extraction of  nuclear matter properties from the $\beta$-equilibrium EoS sets another challenge since the NS interior composition is not known, and even the extraction of the proton fraction puts severe difficulties \cite{Tovar2021,Imam2021,Mondal2021,Essick2021,Essick:2021ezp}. In \cite{Imam2021}, using a Bayesian approach, the authors showed that they were not able to recover the  nuclear matter parameters from the equation of state  of neutron star matter. Also in \cite{Mondal2021} the authors have shown that it was impossible to determine the NS composition  from the $\beta$-equilibrium EoS because  there are multiple solutions. The determination of the symmetry energy from the $\beta$-equlibrium EoS requires the knowledge of the symmetric nuclear matter EoS \cite{Essick:2021ezp}. However,  the uncertainty on the symmetric nuclear matter EoS reflects itself in the precision with which the proton fraction in the NS interior can be determined \cite{Tovar2021}.\\

Machine learning methods have been applied in determining the non-linear map between the $M(R)$ curve and the corresponding $\beta$-equilibrium EoS of NS matter, see \cite{Fujimoto:2017cdo,Fujimoto:2019hxv,Fujimoto:2021zas,Ferreira2019,Morawski:2020izm,Krastev:2021reh,Soma:2022qnv}. In the present study, we will explore DNN to map the properties of nuclear matter from the $\beta$-equilibrium EoS. 
The model is trained on a dataset made of different realizations of $\beta$-equilibrium NS matter and the corresponding nuclear matter properties.
The dataset was generated assuming a quadratic dependence on the isospin asymmetry for the energy per particle of homogeneous nuclear matter, and the iso-scalar and iso-vector parts were Taylor expanded up to fourth order around saturation density. Although, these two assumptions are limitations, they allow us to test the concept.
Finally, we analyze a real case scenario, where 
we apply the trained model to
33 $\beta$-equilibrium EoS from Skyrme and relativistic mean-field (RMF) models \cite{Fortin16} and infer the corresponding  nuclear matter properties. 
The present DNN model allows for the instantaneous inference of the nuclear matter properties once we have a candidate, or a set of candidates, for the true EoS of NS matter.\\

The paper is organized as follows. The EoS parametrization and the method for generating the EoSs are presented in Sec. \ref{sec2}.  The EoS dataset is built in Sec. \ref{sec3} and the DNN model applied in the study is presented in Sec. \ref{sec4}. The results are discussed in Sec. \ref{sec5}, where we analyze the properties of the different EoS data sets and perform a probabilistic inference on the high density region of the EoS. 
Lastly, the conclusions are drawn in Sec. \ref{sec:conclusions}.

\section{Formalism\label{sec2}}

Our goal is to relate the EoS of $\beta$-equilibrium NS matter with the corresponding nuclear matter properties. A great amount of work has been developed in constraining the EoS of NSs using all available information, i.e, astrophysical observations as discussed in the Introduction, terrestrial experiments  (see \cite{OertelRMP16} for a review), ab-initio calculations \cite{Hebeler:2009iv,Gandolfi:2011xu,Hebeler2013,Gezerlis:2013ipa,Drischler:2017wtt}. To constraint the thermodynamic properties of NS matter, a specific EoS parametrization is employed, as generic and flexible as possible, and the parameters are normally estimated from a Bayesian inference framework \cite{Xie:2019sqb,Traversi:2020aaa,Char:2020utj,Malik:2022jqc,Pfaff:2021kse,Malik:2022zol,Li2021a}. However, we ultimately would like to learn not only about the thermodynamic properties of NS matter but we would also like to map that information into the possible degrees of freedom involved and the properties of nuclear matter. \\

We assume that the energy per particle of asymmetric nuclear matter, $E_{\text{nuc}}/A$ , has a quadratic dependence on the isospin asymmetry $\delta =(n _{n}-n _{p})/(n _{n}+n _{p})$ \cite{Bombaci1991,Vidana2009}, 
\begin{equation}{
\frac{E_{\text{nuc}}}{A}\left ( n,\delta  \right )= \frac{E_{\text{SNM}}}{A}\left ( n \right )+S\left ( n \right )\delta ^{2},} 
\label{EoS}
\end{equation}
where $n_n$ and $n_p$ are, respectively, the neutron and the proton densities, $n=n_n+n_p$ is the baryonic density,  ${E_{\text{SNM}}}/{A}$ is the symmetric nuclear matter energy per particle, and $S\left( n \right)$ is the symmetry energy,
\begin{equation}
S\left ( n \right )=\frac{1}{2} \left.\frac{\partial^2 {E_{\text{nuc}}/A}}{\partial \delta^2}\right|_{\delta=0}.
\end{equation}
We perform a Taylor expansion for both the symmetric nuclear matter energy per particle  ${E_{\text{SNM}}}/{A}$ and the symmetry energy $S(n)$ around the saturation density ($n_0$):
\begin{equation}
\frac{E_{\text{SNM}}}{A}\left ( n \right )=E_{\text{\text{\text{sat}}}}+\frac{K_{\text{\text{\text{sat}}}}}{2}\eta^{2}+\frac{J_{\text{\text{\text{sat}}}}}{3!}\eta^{3}+\frac{Z_{\text{\text{\text{sat}}}}}{4!}\eta^{4}, 
\label{snm}
\end{equation}
and
\begin{equation}
S\left ( n \right )=E_{\text{\text{sym}}}+L_{\text{\text{sym}}}\eta+\frac{K_{\text{\text{sym}}}}{2}\eta^{2}+\frac{J_{\text{\text{sym}}}}{3!}\eta^{3}+\frac{Z_{\text{\text{sym}}}}{4!}\eta^{4},
\label{esym}
\end{equation}
where $\eta=({n-n_{0}})/({3n_{0}})$.
The coefficients in Eq.~(\ref{snm}) correspond to,
respectively, the energy per particle $E_{\text{\text{\text{sat}}}}$, the incompressibility $K_{\text{\text{\text{sat}}}}$, the skewness $J_{\text{\text{\text{sat}}}}$, and the kurtosis $Z_{\text{\text{\text{sat}}}}$ at saturation density. 
The expansion coefficients  of symmetry energy $S(n)$ in Eq.~(\ref{esym}) are identified, respectively, as  the symmetry energy  $E_{\text{\text{sym}}}$, and its slope $L_{\text{\text{sym}}}$, curvature $K_{\text{\text{sym}}}$, skewness  $J_{\text{\text{sym}}}$, and kurtosis $Z_{\text{\text{sym}}}$ at saturation density. \\

Nuclear matter inside NSs is in $\beta$-equilibrium, and we denominate the $\beta$-equilibrium energy density by $\varepsilon_\beta$. The relation between $\varepsilon_\beta$ and the nuclear matter energy per particle $E_{\text{nuc}}/A$ is given by
\begin{equation}
\varepsilon_\beta=n \left(\frac{E_{\text{nuc}}}{A}(n,\delta)+\bar{m}_N\right)+\varepsilon_{lep}(n,\delta), 
\label{ebeta}
\end{equation}
where $\varepsilon_{lep}$ designates the leptonic energy density, which for cold-catalysed matter includes the contribution of electrons and muons
$ \varepsilon_{lep}(n,\delta)= \varepsilon_{e}(n,\delta)+\varepsilon_{\mu}(n,\delta)$,
and  
$\bar{m}_N=m_n(1-x)+m_p x$, with $x=n_p/n$ the proton fraction,  which reduces to the nucleon vacuum mass $m_N$ if  $m_n$ and $m_p$ are taken equal to the average nucleon mass $(m_n+m_p)/2$ (see \cite{deTovar:2021sjo}).
The different particle fractions result form the charge neutrality condition, which establishes a relation between $n_p$, $n_e$ and $n_\mu$. Furthermore, $\beta$-equilibrium imposes the following relations between the chemical potentials of neutrons, protons, electrons and muons:
$\mu _{n}=\mu _{p}+\mu _{e}$, $\mu_\mu=\mu_e$. 

\section{Dataset \label{sec3}}
The dataset was generated by considering a quadratic dependence on the isospin asymmetry for the
energy per particle of asymmetric nuclear matter (see Eq.~(\ref{EoS})). Furthermore, both $E_{\text{SNM}}(n)$ and $S\left ( n \right )$ were approximated by their Taylor expansions around saturation density, Eqs.~(\ref{snm}) and ~(\ref{esym}), respectively.
Therefore, an EoS is generated by selecting a set of values (10 in total) for the expansion terms: i) $E_{\text{\text{\text{sat}}}}$, $K_{\text{\text{\text{sat}}}}$, $J_{\text{\text{\text{sat}}}}$, $Z_{\text{\text{sat}}}$ for $E_{\text{SNM}}$; ii) $E_{\text{sym}}$, $L_{\text{sym}}$, $K_{\text{sym}}$, $J_{\text{sym}}$, $Z_{\text{sym}}$ for $S(n)$; and iii) the saturation density $n_0$. Then, the $\beta-$equilibrium EoS is determined from Eq.~(\ref{ebeta}).
Any thermodynamic unstable EoS generated is discarded.\\

A grid of possible values and ranges for the expansion parameters and the saturation density is shown in Table \ref{tab:grid_values}. We use $N=6$ for nine parameters and $N=16$ for the baryonic density of equally separated points within the corresponding ranges for computationally reasons: with this choice the total number of grid points is $16\times 6^{9}$. However, the dataset size is considerable smaller because some points in the grid give rise to thermodynamic unstable EoS. The energy density as a function of the baryonic density, $\varepsilon_{\beta}(n)$, is represented and saved in a vector format: $\pr{\varepsilon_{\beta}(n_{\text{min}}), \varepsilon_{\beta}(n_{\text{min}}+\Delta n),\cdots,\varepsilon_{\beta}(n_{\text{max}})}$, where $n_{\text{min}}=0.08$ fm$^{-3}$,  $n_{\text{max}}=0.3$ fm$^{-3}$, and $\Delta n=0.0055$ fm$^{-3}$, i.e.,
each $\beta$-equilibirum EoS is represented by a vector with $41$ elements. The accuracy of the model was seen to decrease for lower input sizes. The value $41$ for the input size was chosen as a tradeoff between accuracy and the total size of the dataset (and corresponding training time). 
The values for $n_{\text{min}}$ and $n_{\text{max}}$ were chosen by considering densities  close to the saturation density. It was shown in \cite{deTovar:2021sjo} that  the EoS of a given RMF model is well reproduced by a Taylor expansion around saturation density considering terms until the fourth order. Moreover, the coefficients of the lower order terms, in particular, until the second order are very close to the corresponding properties of the RMF model: the coefficients of the quadratic terms are reproduced within less than 5\% and the coefficients of the lowest order terms  within $\lesssim 0.5\%$. The number of points chosen to represent each EoS (i.e., 41) was considered large enough to characterize the EoS in the range $0.08\leqslant n \leqslant 0.30$  fm$^{-3}$ and still acceptable from the computational necessities.  The lower limit was defined at approximately above the crust-core transition, and, therefore, avoids the region of the EoS that includes clusterization. The upper limit was chosen to avoid the possible onset on non-nucleonic degrees of freedom.  However, it should be pointed out that in some studies non-nucleonic degrees of freedom occur below twice saturation density as  $\Delta$-baryons \cite{Li:2018jvz,Raduta:2021xiz,Marquez:2022gmu}  or quark matter \cite{Ferreira:2020kvu}.\\

The final dataset is composed of $89567940$ EoS in $\beta-$equilibrium and the corresponding nuclear matter properties. 
A random split of the dataset into $90\%/10\%$ was done, corresponding to the train/test sets.
The test set contains a total number of 8956794 EoS which will
allow us to measure the real (out-of-sample) accuracy of the ML model.
The use of a fixed grid with equally spaced points was to access the model metrics in a case where the model's interpolation is uniform across the whole space of parameters. A possible alternative consists in sampling points from the 10-dimensional space of parameters using a multivariate Gaussian distribution, with the present mean and variance values for each parameter given in Table \ref{tab:grid_values}. This would increase the model accuracy as most of the nuclear models, used in the last part of work, are cluster around these mean values.  
\begin{table}[htb]
\begin{tabular}{c|c|c}
 & $\pr{\text{min},\text{max}}$ &\# of points\\
 \hline
$E_{\text{\text{sat}}}$ & $[-16.7,-14.9]$ & $6$\\
\hline
$K_{\text{\text{sat}}}$ &  $[170,290]$ & $6$\\
\hline
$J_{\text{\text{sat}}}$ & $[-900,150]$ & $6$\\
\hline
$Z_{\text{\text{sat}}}$ & $[-3500,2500]$ & $6$\\
\hline
$E_{\text{sym}}$& $[26,38]$ & $6$\\
\hline
$L_{\text{sym}}$&  $[15,105]$ & $6$\\
\hline
$K_{\text{sym}}$&  $[-400,200]$ & $6$\\
\hline
$J_{\text{sym}}$& $[-1200,1200]$ & $6$\\
\hline
$Z_{\text{sym}}$& $[-3500,2500]$ & $6$\\
\hline
$n_0$ & $[0.145,0.166]$ & 16\\
\end{tabular}
\caption{Range of values for the parameters and the number of equally separated points.}
\label{tab:grid_values}
\end{table}

\section{Deep Neural Network model \label{sec4}}
The present problem is a multivariate regression (supervised) task. Furthermore, we are concerned with the interpolation capacity, and not the extrapolation, of the model  as the grid of parameters is wide enough to accommodate almost all nuclear models.  
Neural networks is a powerful supervised learning algorithm that approximates the function represented by the data. Neural networks is a perfect framework 
as it can approximate any function (universal approximation theorem) with arbitrary precision. 
Our goal is to learn the map
$
f: \mathbf{X} \longrightarrow \mathbf{Y},
$
which, in our case, $f: \mathbb{R}^{41} \longrightarrow \mathbb{R}^{10}$, i.e., the input space represents the EoS in $\beta-$equilibrium, a vector of length 41 with $\varepsilon_{\beta}(n_i)$ values, and the output space represents the corresponding nuclear matter properties of the EoS, a vector of 10 elements. \\

We have considered two different model structures, one that predicts the entire 10 properties of nuclear matter 
and another that ignores the two highest order terms, $Z_{\text{sym}}$ and $Z_{\text{\text{sat}}}$, and thus the model has an output size of length $8$, i.e., $f: \mathbb{R}^{41} \longrightarrow \mathbb{R}^{8}$. After extensive analyzis, and considering the real applicability of the model, we concluded that predicting the 8 lowest order parameters has several advantages. Although is has no clear advantage from the training/test point of view using the present dataset, apart from becoming a smaller model and a faster training stage, there is a crucial point that distinguishes a real scenario from a EoS of the dataset. A data point in the dataset is an approximation (Taylor expansion around saturation density) of a real EoS in the range $0.08\leqslant n \leqslant 0.30$  fm$^{-3}$, and a real EoS has in principle a infinite number of terms in the expansion.
The higher order terms of the approximation will be thus effective terms (see \cite{deTovar:2021sjo}), deviating from the real values as they are taking into consideration  higher order terms from the full EoS. We concluded that being effective, not reflecting the real values of the EoS, it is better to exclude them from the map. In other words, this is a way of informing the DNN that, given a real EoS, we want to extract the lowest order terms regardless of the effective higher order terms $Z_{\text{sym}}$ and $Z_{\text{\text{sat}}}$.
\subsection{Structure}
The present task is a multivariate regression problem (supervised learning), i.e., our data points are represented by $(\mathbf{y_i},\mathbf{x_i})$ and we want to learn the map $\mathbf{y}(\mathbf{x})$. In our case, the input space is $\mathbf{x_i} \in \mathbb{R}^{41}$ while the output space $\mathbf{y_i} \in \mathbb{R}^{8}$, where $\mathbf{x_i}=\pr{\varepsilon_{\beta}(n_1),\varepsilon_{\beta}(n_2),\cdots,\varepsilon_{\beta}(n_{41})}$ denotes the energy density of $\beta-$equilibrium matter, and $\mathbf{y_i}=\pc{E_{\text{\text{sat}}},K_{\text{\text{sat}}}, J_{\text{\text{sat}}}, E_{\text{sym}},L_{\text{sym}}, K_{\text{sym}}, J_{\text{sym}}, n_0}$ the corresponding set of parameters. 

\begin{figure}[htb]
\includegraphics[width=.95\linewidth]{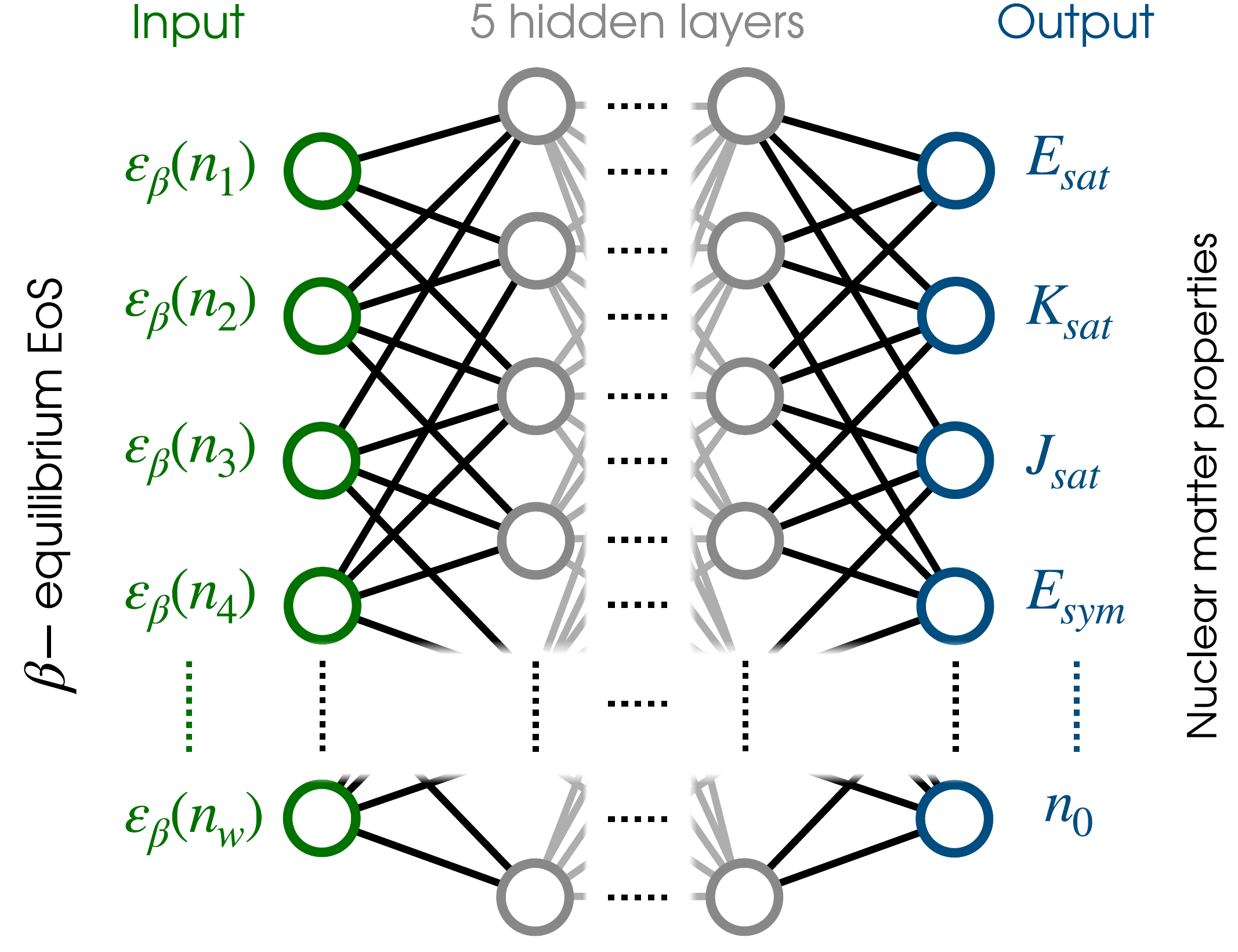} 	
\caption{Schematic and simplified representation of the DNN model, where only a fraction of the connections between nodes is shown. A rigorous representation would display all connections (weights) between any neuron in a given layer and all neurons of the previous layer, i.e., a densely connected neural network.  
}
\label{fig:dnn_structure}
\end{figure}

We use a densely connected feed-forward DNN machine learning model. After exploring several architectures for the DNN, we have selected the one shown in Table \ref{tab:model}. The model contains 5 hidden layers and the sigmoid function, $\sigma(z)=1/(1+e^{-z})$, for the activation functions (Fig.~\ref{fig:dnn_structure} shows a schematic representation of the model). The characteristics of the DNN structure, which defines the final model, were selected using a cross-validation procedure that we explain in the next section.

\begin{table}[htb]
\begin{tabular}{c|c|c}
Layer & Activation function & size \\
 \hline
0  & - & 41 \\
1 & Sigmoid & 80 \\
2 & Sigmoid & 80 \\
3 & Sigmoid & 80 \\
4 & Sigmoid & 40 \\
5 & Sigmoid & 15 \\
6 & Linear & 8 \\
\hline
\end{tabular}
\caption{Structure of the densely connected feed-forward
DNN model.}
\label{tab:model}
\end{table}

\subsection{Training}
Training the DNN model consists in adjusting the weights of the model so that a target loss function is minimized. 
We have used the mean squared error (MSE) for the loss function,
$$
L(\mathbf{w})=\frac{1}{M}\sum_{i=1}^M(\hat{\mathbf{y}}_i(\mathbf{w})-\mathbf{y}_i)^2,
$$
where $\hat{\mathbf{y}}_i(\mathbf{w})$ are the DNN model predictions ($\mathbf{w}$ denotes the DNN model's weights), $\mathbf{y}_i$ are the real values, and $M$ is the batch size. The optimization problem is solved using the Adam algorithm \cite{kingma2014adam} (with a learning rate of $0.001$), which is a stochastic gradient descent method that relies on adaptive estimation of first- and second-order moments, with a batch size of $5120$. A cross-validation with a split fraction of $0.2$ was used to have an out-of-sample evaluation of models. We have trained for a total of 4000 epochs (no gain was seen for a larger number of epochs). The size of the dataset made a possible extensive grid search on all the model hyperparameters computationally unfeasible. However, we tried tens of structures with different combinations of the number of layers, number of neurons per layer, activation functions, and learning rate. The best model we obtained is described in Table \ref{tab:model}.

\subsection{Test}

After the training stage is completed and the best model is reached (Table \ref{tab:model}), we access the final performance on the test set (data neither used for training nor for validation), composed of 8956794 EoS. The standard deviation of the residuals, $\sigma_{\bm{\epsilon}_i}=\sqrt{\text{Var}(\bm{\epsilon}_i)}$, where $\bm{\epsilon}_i=\bm{A}_i^{\text{model}}-\bm{A}_i$ is the model's residuals vector for the quantity $A_i$, which designates one of the possible 8 nuclear matters properties,
and Var is the variance operation. The results are presented in Table \ref{tab:test} and they show that a high accuracy was attained by the DNN model.
The lower order terms were extracted with higher accuracy due to the smaller range of possible values, compared with the higher order ones, and thus better interpolation precision.

\begin{table}[htb]
\begin{tabular}{c|cc}
 $A_i$& $\overline{\bm{\epsilon}_i}$& $\sigma_{\bm{\epsilon}_i}$\\
  \hline
$n_0$&-0.0000	& 0.0010 	\\
$E_{\text{\text{sat}}}$ &	-0.0210	&	0.0899	 \\
$K_{\text{\text{sat}}}$ &  1.3517& 10.9070\\
$J_{\text{\text{sat}}}$ &  -9.3079& 	137.2444\\
$E_{\text{sym}}$& 	-0.0114& 	0.2036	\\
$L_{\text{sym}}$& 	-0.0123	 & 0.9798	\\
$K_{\text{sym}}$& -2.8549&15.5004 \\
$J_{\text{sym}}$& 5.7081& 167.1366\\
 \hline
\end{tabular}
\caption{Mean and standard deviation, $\sigma_{\bm{\epsilon}_i}=\sqrt{\text{Var}(\bm{\epsilon}_i)}$,  of the model residuals,
$\bm{\epsilon}_i=\bm{A}_i^{\text{model}}-\bm{A}_i$, determined on the test set. The saturation density $n_0$ is given in fm$^{-3}$ and all other quantities in MeV.}
\label{tab:test}
\end{table}

\section{Nuclear models application \label{sec5}}
Having shown that the DNN model was able to predict the test set with high accuracy, we then apply it to a more general EoS.
To simulate a real application of the model, i.e., to extract the nuclear matter properties directly from the equation of state of NS matter, we apply the model to a set of nuclear models.  The extraction of information from the EoS is ideally reached via some numerical inference procedure that uses all available NS data.
As mock EoS, we consider a set of 33 unified EoS built from a relativistic mean field (RMF) approach and non-relativistic Skyrme interactions \cite{Fortin:2016hny}. \\

Before we discuss the results, let us stress again the complexity of the present inference task and our model assumptions and limitations. Our ML model was trained in a dataset that assumes: i) the EoS of nuclear matter depends on the matter isospin asymmetry via a quadratic dependence only; ii) the Taylor expansions, Eqs~(\ref{snm}) and ~(\ref{esym}), are good approximations for the nuclear matter EoS in the density range $0.08\leqslant n \leqslant 0.30$  fm$^{-3}$; iii) in the range $0.08\leqslant n \leqslant 0.30$  fm$^{-3}$, NS matter is composed of nucleons, electrons, and muons in $\beta$-equilibrium. Any deviation from these assumptions will generate model prediction errors.
Furthermore, it should be noticed that the training set covers a parameter space defined in Table \ref{tab:grid_values}. While the range of nuclear matter parameters is well within the expected values (see \cite{Margueron:2017eqc}), the DNN model would rely on its extrapolation capacity if we input a $\beta-$equilibrium EOS with nuclear matter parameters outside that range. Some nuclear models used as input in this section are such examples (NL3 with $L_{\text{sym}}=118$ MeV, GM1 with $K_{\text{sat}}=300$ MeV, DDH$\delta$ with $E_{\text{sym}}=25$ MeV). \\

\begin{figure*}[htb]
\includegraphics[width=.95\linewidth]{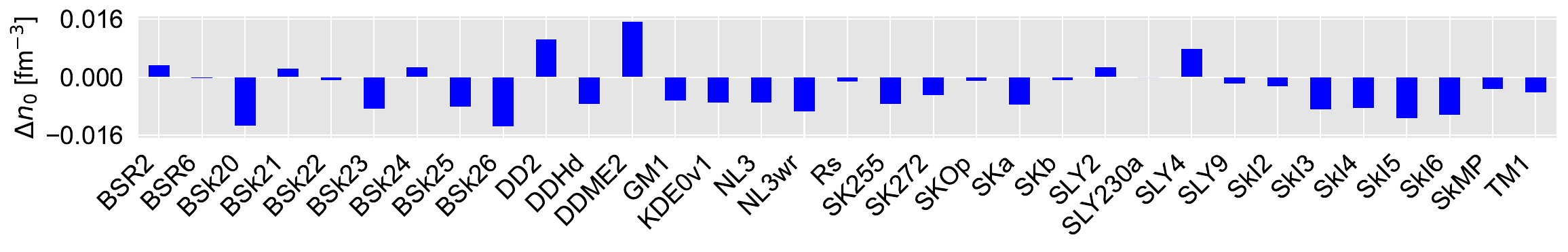} \\
\includegraphics[width=.95\linewidth]{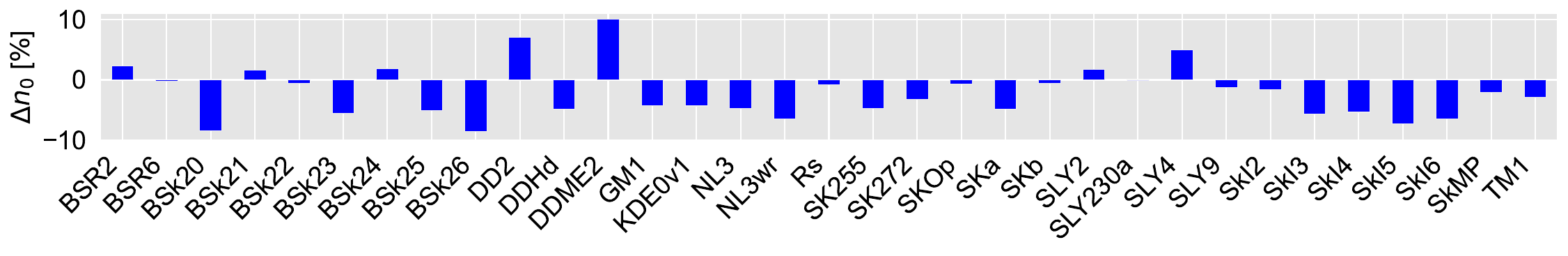} 	
\caption{Model residuals (top) and percentage (bottom) for the saturation density $n_0$.}
\label{fig:density}
\end{figure*}

\begin{figure*}[htb]
\includegraphics[width=.95\linewidth]{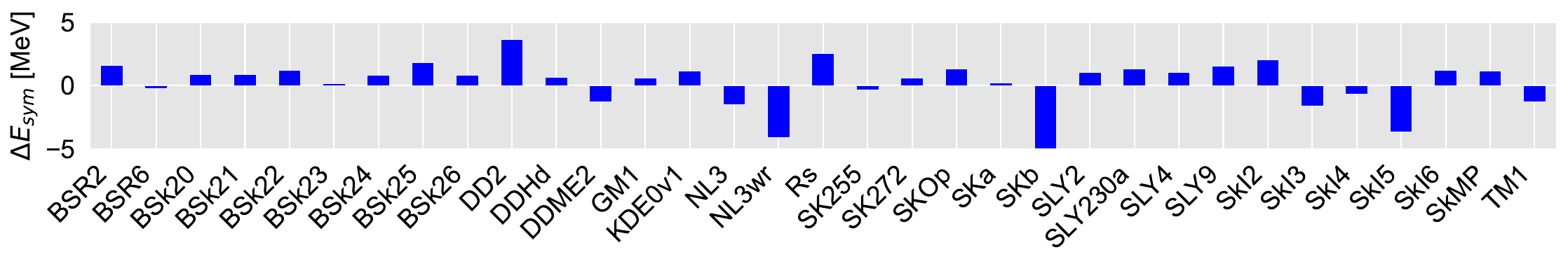} \\
\includegraphics[width=.95\linewidth]{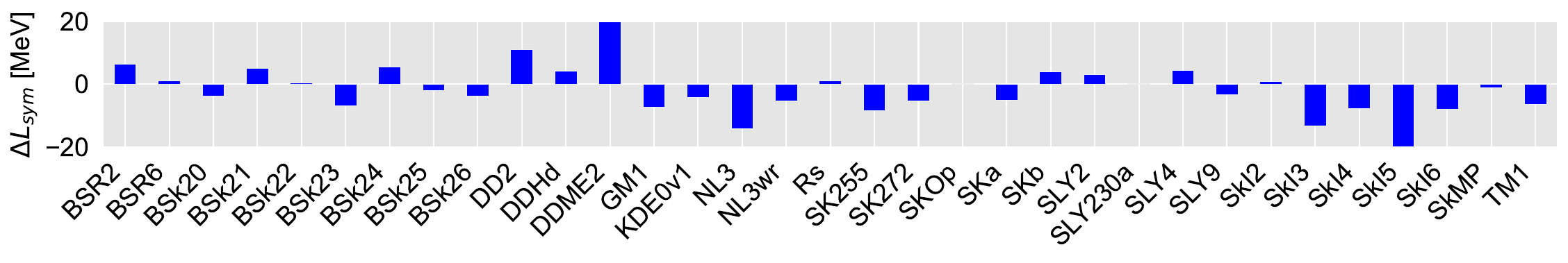} \\
\includegraphics[width=.95\linewidth]{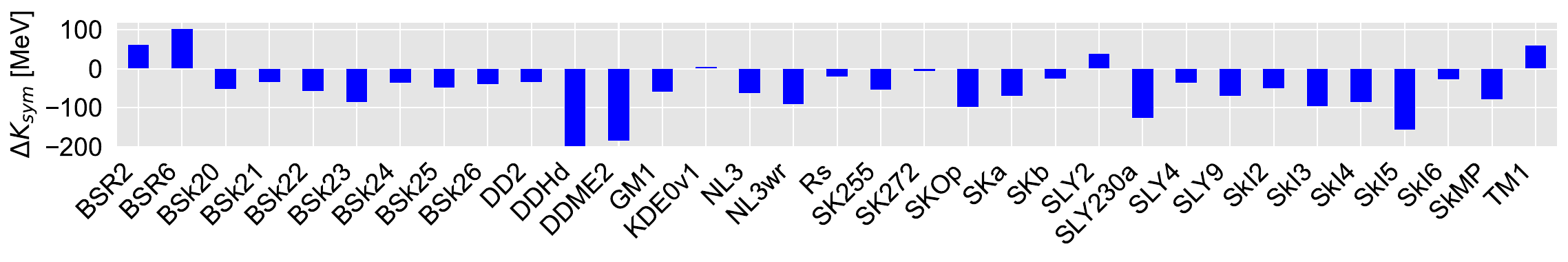}
\caption{Model residuals for the iso-vector properties, , $E_{sym}$ (top), $L_{sym}$ (middle)  and $K_{sym}$ (bottom).}
\label{fig:isovector}
\end{figure*}

\begin{figure*}[htb]
\includegraphics[width=.95\linewidth]{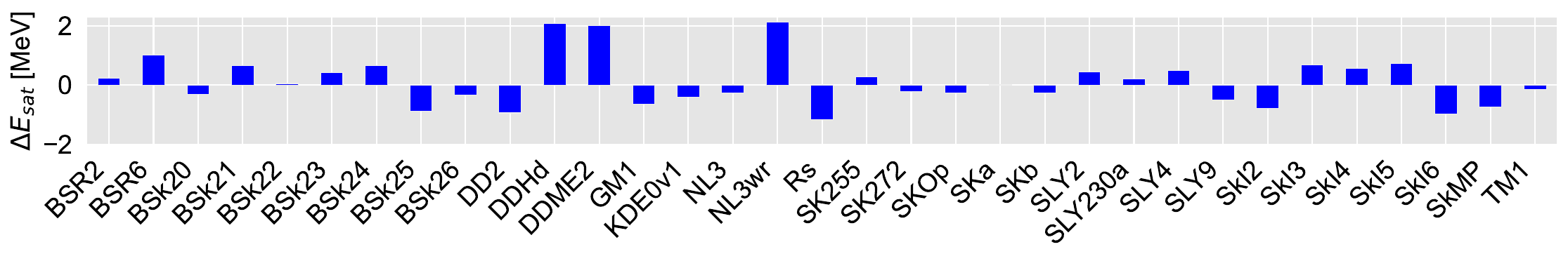} \\
\includegraphics[width=.95\linewidth]{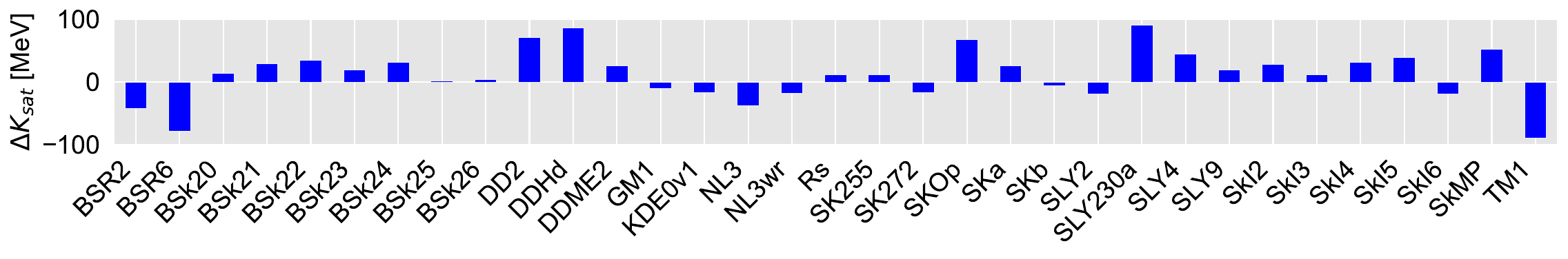}
\caption{Model residuals for the iso-scalar properties, $E_{sat}$ (top) and $K_{sat}$ (bottom).}
\label{fig:isoscalar}
\end{figure*}

The model residuals are shown, for each nuclear model, in Figs~\ref{fig:density}, \ref{fig:isovector}, and \ref{fig:isoscalar}.  The residuals for $J_{\text{\text{sat}}}$ and $J_{\text{sym}}$ are not computed as they were not determined in \cite{Fortin:2016hny}. Besides, as already mentioned, taking a Taylor expansion until fourth order, the coefficients above the second order are expected to be effective, because they have to take into account the effect of the missing terms \cite{deTovar:2021sjo}. Therefore, a deviation from the true values is expected for the higher order terms. To complement the figures,
we present in Table~\ref{tab:final_test} some  statistics involving the residuals, that summarize the uncertainty estimation associated with the nuclear matter parameters extracted from a real $\beta$-equilibrium NS EoS. 
These uncertainties are smaller than the ones obtained for each property from a compilation of experimental data in \cite{Margueron:2017eqc}, where an overview of the present uncertainty on the different nuclear empirical parameters can be found. For instance, the standard deviation of the model's deviations for $L_{\text{sym}}$ and $K_{\text{sat}}$ are, respectively, $12.85$ MeV and $41.02$ MeV, which are smaller than current uncertainties \cite{Margueron:2017eqc}. This shows that the DNN model is an efficient and reliable tool for extracting the lowest order parameters.  As expected, it is the second order parameters that have a larger uncertainty associated.  An improvement in the results would occur if higher order terms are considered in expansions defined in Eqs.~(\ref{snm}) and (\ref{esym}) and if terms beyond quadratic are considered in the isospin asymmetry. In particular, some information on this point will obtained analysing the models that present larger deviations, e.g., some RMF models. Let us also comment that if the lowest orders are obtained with large deviations, the following orders will be affected. In fact,  it is very important to be able to determine the saturation density with a large accuracy and this explains why a larger number of grid points was considered for this quantity. Since the higher orders are affected by the accuracy of the lowest ones, this may indicate that also the lowest order properties $E_{\text{\text{sat}}}$ and $E_{\text{sym}}$ should  also be considered with a larger number of grid points (see Table \ref{tab:grid_values}). 
\\

\begin{figure}[htb]
\includegraphics[width=.95\linewidth]{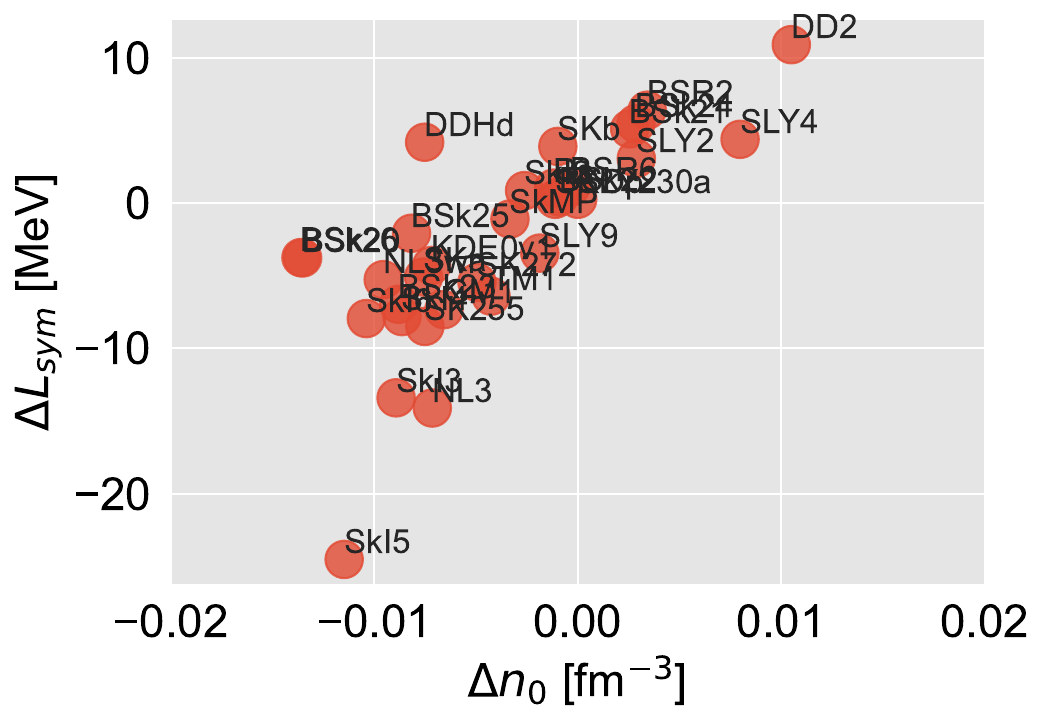}\\ 
\includegraphics[width=.96\linewidth]{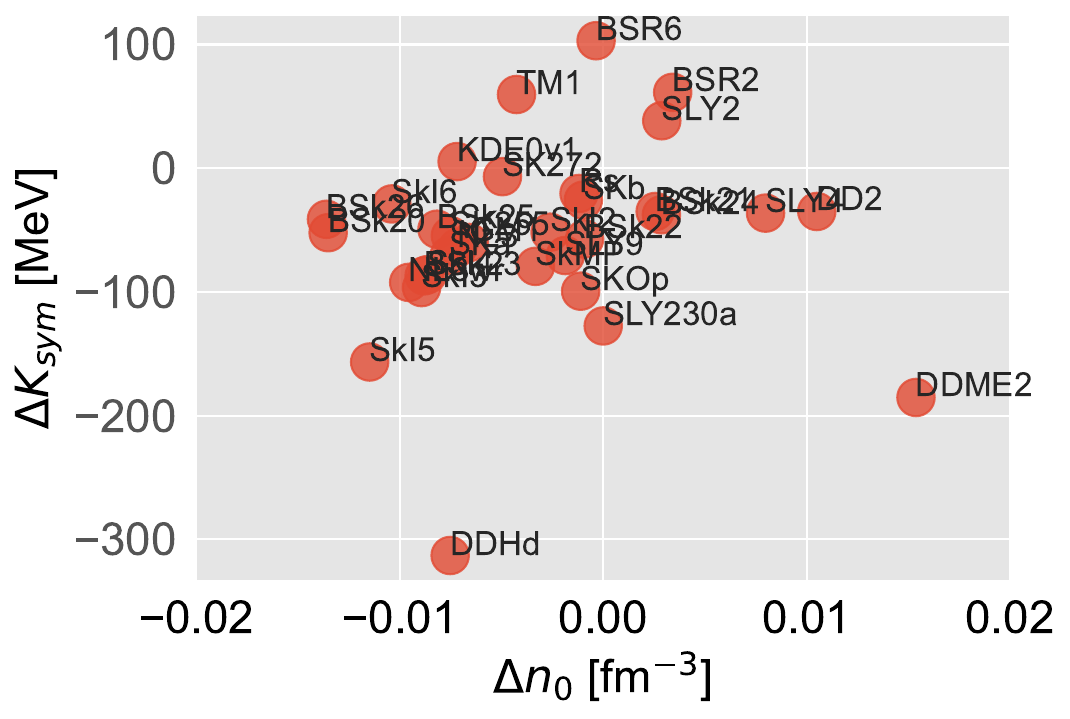} \\
\includegraphics[width=.96\linewidth]{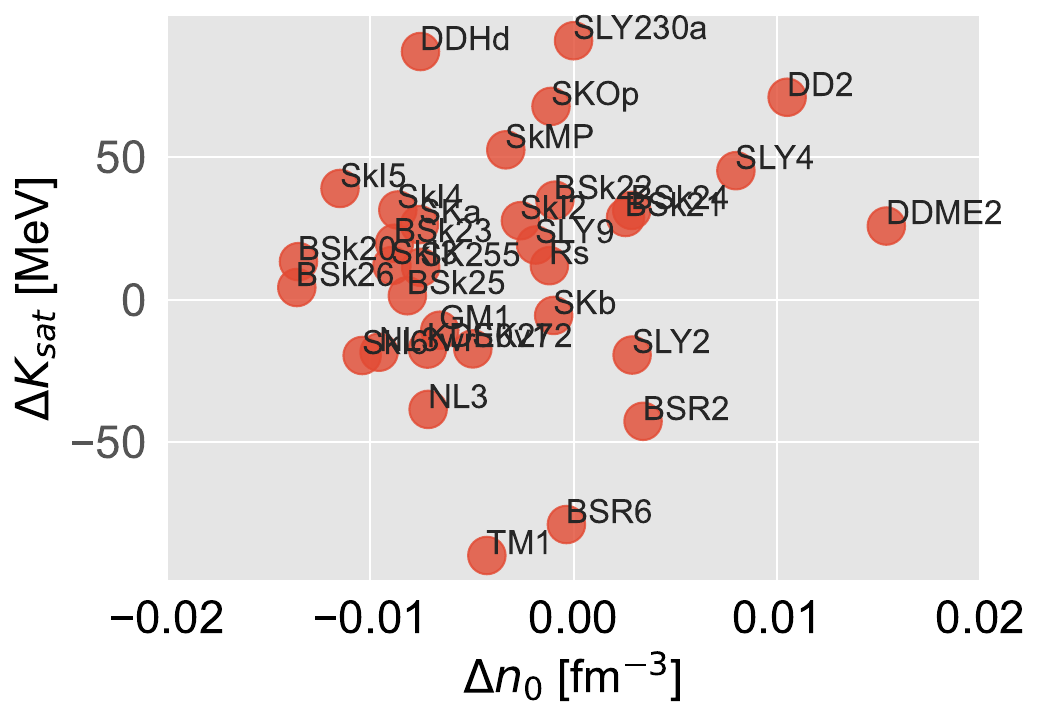}
\caption{Scatter plot of the model residuals $\Delta L_{\text{sym}}$ vs. $\Delta n_{0}$ (top), $\Delta K_{\text{sym}}$ vs. $\Delta n_{0}$ (middle), and $\Delta K_{\text{sat}}$ vs. $\Delta n_{0}$ (bottom). To make the top plot easier to read, the DDME2 residual $(\Delta L_{\text{sym}},\Delta n_{0})=(59.62872 \text{ MeV},0.0153 \text{ fm}^{-3})$ is not shown.
}
\label{fig:scatter}
\end{figure}

Figure \ref{fig:scatter} shows three scatter plots of
$L_{\text{sym}}$ (top), $K_{\text{sym}}$ (middle), and $K_{\text{\text{sat}}}$ (bottom) residuals as a function of 
the $n_0$ residuals. The clear  positive correlation between $\Delta L_{\text{sym}}$ and $\Delta n_{0}$  indicates that a greater accuracy in $L_{\text{sym}}$ depends on the accuracy of the saturation density prediction. This pattern is present but with a smaller  correlation  in $\Delta K_{\text{sym}}$ vs $\Delta n_{0}$, showing that a better performance is attainable for the symmetry energy  increasing the number of $n_0$ points in the dataset, and thus increasing the interpolation accuracy.  

\begin{table}[htb]
\begin{tabular}{c|cccc}
 $A_i$& $\overline{\bm{\epsilon}_i}$& 
 $\sigma_{\bm{\epsilon}_i}$&
$10\%$& $90\%$\\
  \hline
 $n_0$    &  -0.0034	 & 0.0067&	-0.0102&0.0033\\
$E_{\text{sat}}$ & 0.1172& 0.8491&	-0.8636&0.9528\\
$K_{\text{sat}}$ &11.9596 & 41.0238&	-34.6530	&64.6642\\
$E_{\text{sym}}$& 0.1693& 2.1487&	-1.5729	&1.7502\\
$L_{\text{sym}}$& -0.7340& 12.8528&-8.3805	&5.3971\\
$K_{\text{sym}}$& -56.1287& 74.6052&-121.8530	&31.7860\\
 \hline
\end{tabular}
\caption{Mean, the standard deviation, 
and the 10th and 90th percentiles of the model residuals,
$\bm{\epsilon}_i=\bm{A}_i^{\text{model}}-\bm{A}_i$, determined for 33 nuclear models. The saturation density $n_0$,  is given in fm$^{-3}$ and all other quantities in MeV.}
\label{tab:final_test}
\end{table}

\section{Conclusions}
\label{sec:conclusions}
With the  perspective that  in the near future the mass and radius of a large number of NSs will  be determined with a small accuracy, it will  also be possible to extract the NS EoS with a small uncertainty.  This opens a new door in nuclear physics  since inside NSs large baryonic densities  and isospin asymmetries are expected, not attainable in the laboratory. The question that remains is how to extract nuclear matter properties from the knowledge of a NS EoS. In the present work, DNN were explored to predict nuclear matter properties from the knowledge of the $\beta$-equilibrium EoS. For training, we have generated a large dataset of $\beta$-equilibrium EoS from the corresponding nuclear matter properties. The final model, selected from a cross-validation procedure, show high  accuracy on the test set. 

The DNN model was then applied to a real case scenario, 
where the nuclear properties of 33 nuclear models, obtained within a RMF approach or a Skyrme force description, where extracted. The DNN was, on average, quite successful and the residuals for the nuclear matter properties are smaller than the estimated uncertainties obtained from experimental data on the same properties. In future work, the EoS for which the DNN did not perform so well will be analysed in order to identify possible improvements, which may include a redefinition of the number of grid points included to generate the EoS dataset, the inclusion of higher order terms in the Taylor expansion, or the inclusion of the contributions beyond the parabolic approximation for the isospin asymmetry.  

In future work, we will apply the DNN model to a dataset of density dependent RMF models \cite{Malik:2022zol}, constructed from a Bayesian analysis, to compare the extracted  and the actual distributions of the nuclear matter properties. 
The use of non-parametric representations, such as Gaussian-process (GP) regression, has been explored to generate large datasets of NS matter EoS \cite{Landry:2018prl,Landry:2020vaw,Essick:2019ldf,Gorda:2022jvk}. GP is a probabilistic model that, with a suitable choice of the kernel function, effectively constrains the EoS to be a smooth  function of the baryonic density \cite{Landry:2018prl}. These datasets are generally conditioned on astrophysical observations, low density {\it ab-initio} calculations, and perturbative QCD results, and represent the most likely region for the $P(n)$ of $\beta-$equilibirum NS matter.
However, the extraction of nuclear matter properties from these agnostic descriptions is difficult. We consider that the method presented in this work enables the instantaneous inference of the nuclear matter properties from some of these agnostic descriptions.

The model is available from the corresponding author upon request.

\section{Acknowledgments}

This work was partially supported by national funds from FCT (Fundação para a Ciência e a Tecnologia, I.P, Portugal) under the Projects No. UID/\-FIS/\-04564/\-2019, No. UID/\-04564/\-2020, and No. POCI-01-0145-FEDER-029912 with financial support from Science, Technology and Innovation, 
in its FEDER component, and by the FCT/MCTES budget through national funds (OE). V.C. acknowledges a FCT grant received through Centro de Física da Universidade de Coimbra.

%\bibliography{biblio.bib}


\end{document}